\documentclass[11pt]{article}

\PassOptionsToPackage{table}{xcolor}
\usepackage[preprint]{acl}
\usepackage[table]{xcolor}
\usepackage{xcolor}
\usepackage{amssymb}
\usepackage{listings}
\newcommand{\cmark}{\textcolor{green!55!black}{\checkmark}}
\newcommand{\xmark}{\textcolor{red!75!black}{$\times$}}
\lstdefinestyle{memjson}{basicstyle=\ttfamily\scriptsize,columns=fullflexible,
  breaklines=true,breakatwhitespace=true,numbers=none,xleftmargin=6pt,
  showstringspaces=false,escapeinside={(*}{*)}}

\usepackage{times}
\usepackage{latexsym}

\usepackage[T1]{fontenc}
\usepackage[utf8]{inputenc}
\usepackage{microtype}
\usepackage{inconsolata}
\usepackage{graphicx}
\usepackage{booktabs}
\usepackage{multirow}
\usepackage{amsfonts}
\usepackage{amsmath}
\usepackage{xcolor}
\usepackage{nicefrac}
\usepackage{colortbl,caption}
\usepackage[most]{tcolorbox}
\tcbuselibrary{breakable,skins}

\usepackage{xspace}
\newcommand{\rthree}[0]{\textsc{R$^3$-SQL}\xspace}
\newcommand{\ours}[0]{\textsc{MaP-SQL}\xspace}

\title{Replacing Training with Memory: Listwise Selection for Text-to-SQL}

\author{
    Yeonseok Jeong\textsuperscript{1},
    Soyoung Yoon\textsuperscript{1},
    Seongjun Lee\textsuperscript{2},
    Seung-won Hwang\textsuperscript{1}\thanks{Corresponding Author} \\
    IPAI, Seoul National University\textsuperscript{1},
    KAIST\textsuperscript{2} \\
    \texttt{\{jys3136, soyoung.yoon, seungwonh\}@snu.ac.kr} \\
    \texttt{epaksa@kaist.ac.kr}
}

\begin{document}
\maketitle

\begin{abstract}
Modern Text-to-SQL systems often follow \textit{generate-execute-select} pipelines, generating multiple candidate queries then selecting the best one.
Listwise selection, by jointly comparing multiple candidates, has been widely adopted, but fine-tuning listwise selectors is costly.
We thus propose a fine-tuning-free listwise selector.
We replace two major fine-tuning objectives with inference-time strategies: (1) learning selection criteria as ordering and (2) mitigating positional bias.
First, we build reusable structured memories instead of learning selection behavior as model parameters.
Given a question, \textbf{\ours} retrieves memories distilled from training data that encode how natural language maps to schema elements, SQL operations, and expected outputs. 
These memories serve as explicit decision criteria for evaluating candidates in a listwise manner.
Second, to mitigate ordering bias of listwise selectors, we aggregate rankings across multiple input permutations, with inference cost optimized by execution results and pointwise scoring. 
Our approach improves selection accuracy while maintaining efficiency and compatibility with existing large language models. 
Across Text-to-SQL benchmarks, it produces more stable selection without fine-tuning and fewer unnecessary comparisons than existing methods.
On BIRD-dev, it outperforms the previous state-of-the-art selector-based method \rthree by 2.02 execution accuracy points on average using the same candidate sets, with 2.92× fewer tokens.\footnote{\href{https://github.com/ldilab/MAP-SQL}{https://github.com/ldilab/MAP-SQL}}
\end{abstract}

\section{Introduction}
\label{sec:intro}

\begin{figure}[t]
  \centering
  \includegraphics[width=\linewidth]{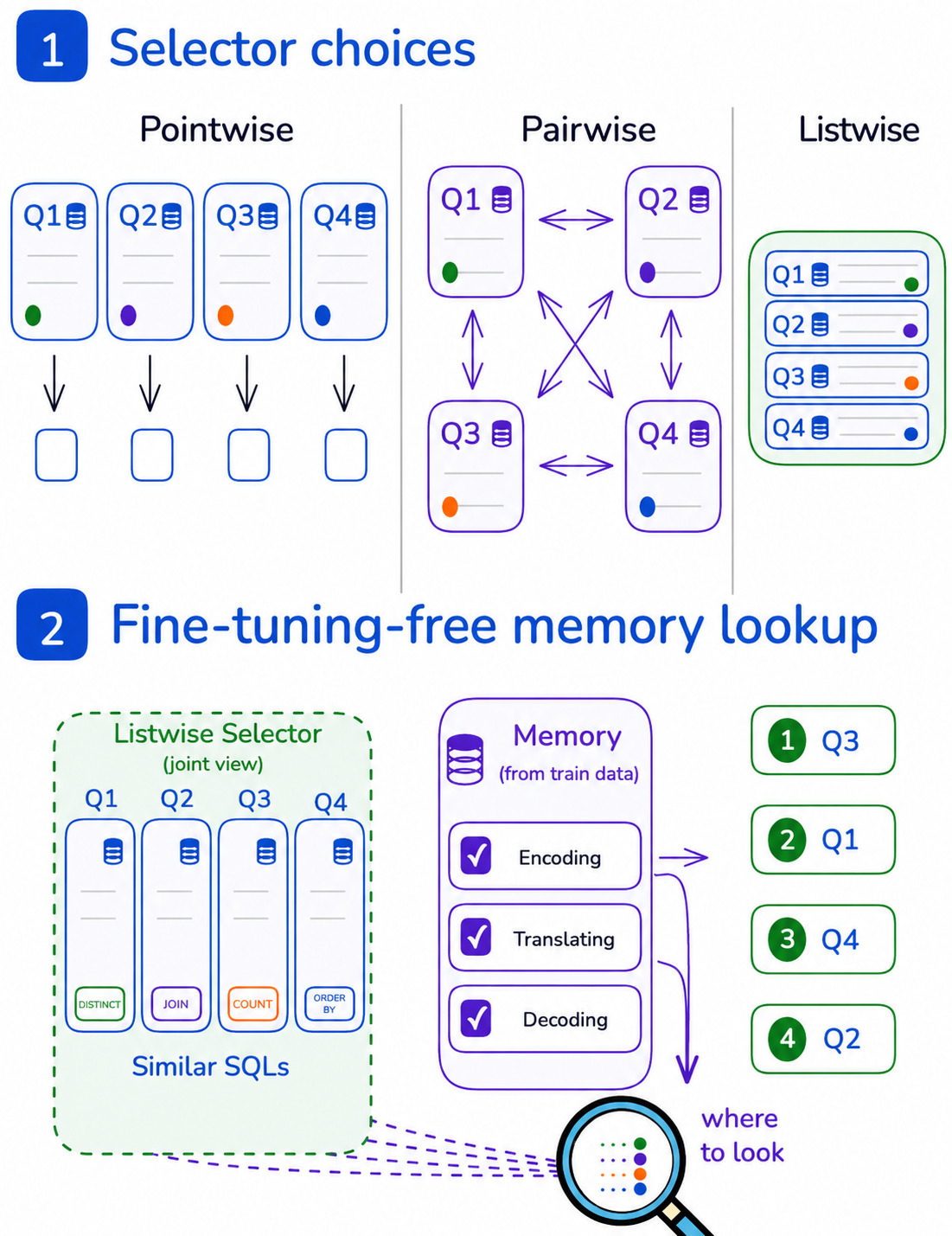}
  \caption{Overview of our proposed SQL selection. Our method uses (1) a listwise selector to jointly compare SQL candidates and (2) fine-tuning-free memory lookup to better capture subtle SQL differences.}
  \label{fig:intro}
\end{figure}

Modern Text-to-SQL systems ~\cite{cscsql, genasql, xiyansql} increasingly follow a \emph{generate--execute--select} pipeline, where multiple SQL candidates are produced and a selector chooses the best one~\cite{li2025omnisql, arctic2025text2sql, wang2025agentar}.
Figure~\ref{fig:intro} illustrates such a pipeline: four candidate queries are generated and then evaluated by a selector.
Depending on how many candidates are evaluated at a time, selectors can be categorized as pointwise~\citep{contextual2025, tritto2025gradesql}, pairwise~\cite{pourreza2025chase, bai2025judgesql}, or listwise.
Pointwise selectors score candidates independently and miss cross-candidate differences, while pairwise selectors compare candidates more directly but require $O(n^2)$ comparisons.
Listwise selection avoids exhaustive quadratic pairwise comparisons by evaluating multiple candidates jointly, allowing it to capture subtle differences that pointwise or pairwise methods may miss~\cite{rankgpt}.

However, this benefit comes at a cost, as fine-tuning listwise selectors is costly. 
Each training example involves multiple SQL queries and their execution results, resulting in long input contexts and high computational overhead. 
As this makes learned listwise selection difficult to scale, our approach replaces the two main roles of fine-tuning with inference-time strategies.

First, fine-tuning provides selection criteria that guide ranking decisions. We replace this by constructing structured memories, as illustrated in Figure~\ref{fig:intro}(2), distilled from training data. Each memory encodes how natural language maps to schema elements, SQL operations, and expected outputs. Given a test question, we retrieve relevant memories and use them as explicit decision criteria for listwise candidate comparison.

Second, fine-tuning aims to mitigate a well-known positional bias in listwise selection, known as ``lost-in-the-middle'' bias~\cite{liu2024lost, tang2024found}. We instead address this at inference time by aggregating rankings across multiple input permutations, with a cost-aware design that leverages execution signals and selective pointwise scoring.

By combining these two components, we present \textbf{M}emory \textbf{a}nd \textbf{P}ermutation for Listwise \textbf{SQL} selection (\textbf{\ours}). 
Our approach eliminates the need for additional fine-tuning while retaining the benefits of joint candidate comparison.
In this paper, \emph{fine-tuning-free} means that \ours does not update selector parameters. It still uses pretrained models and labeled question--SQL pairs to construct retrieval memories.

Our approach is simple, efficient, and compatible with off-the-shelf language models, making it practical for real-world Text-to-SQL systems.
On BIRD-dev~\cite{li2024bird}, Spider-test~\cite{yu2018spider}, and EHRSQL~\cite{lee2022ehrsql}, our method improves over the previous state-of-the-art selector \rthree by 2.02, 0.53, and 0.68 execution accuracy points on average when both methods use the same candidate pools.
Our method also requires 6.54×, 6.85×, and 7.29× fewer selector calls and 2.92×, 2.12×, and 4.16× fewer tokens on the three benchmarks.
\section{Related Work}

We review existing selection paradigms and their training objectives.
Lastly, we present our distinctions of fine-tuning-free approaches.

\label{sec:related}
\subsection{Selector Paradigms in Text-to-SQL}

Recent Text-to-SQL systems include a selection step after generating multiple SQL candidates to improve performance at test time.
Majority voting~\citep{cscsql} selects the most frequent execution outcome, but a larger incorrect group can dominate a smaller correct one.
Pointwise selectors~\citep{contextual2025, tritto2025gradesql} score each candidate independently.
This lacks comparative insights across candidates, leading to inconsistent scoring and suboptimal performance~\cite{long2025precise}.
Pairwise selectors~\citep{pourreza2025chase, wang2025agentar} compare all pairs of candidates within the candidate set.
However, the number of comparisons grows quadratically with the candidate set, making it computationally inefficient.
XiYan-SQL~\cite{xiyansql} adopts a listwise selector that compares all candidates within a window simultaneously.
MCS-SQL~\cite{lee2024mcs} performs multiple-choice selection without selector fine-tuning. It sorts candidates before presenting them to the selector.
In this work, however, the challenges of listwise selection and potential improvements to address them have not been explored.

\subsection{Training for Listwise Reranking and Positional Bias}

Existing listwise selectors require training to encode selection behavior in model parameters.
However, jointly considering a long list of generations requires expensive training on candidate queries and execution results.
In addition, long inputs expose the model to positional bias, commonly known as the ``lost-in-the-middle'' problem~\cite{liu2024lost}.
\rthree \cite{r3-sql} addresses positional bias in training by using a pointwise selector as a tie-breaker when the pairwise selector cannot confidently distinguish between candidates.
Delaying bias mitigation to inference time through self-consistency can incur up to $O(n!)$ calls in principle~\cite{tang2024found, zeng2024llm}. Although optimization strategies for relevance ranking have been studied~\cite{lee2025inference}, no such work has addressed our target problem.

\subsection{Our Distinction: Inference-time Strategies}

We identify two key roles of fine-tuning: fine-tuning selection criteria and mitigating positional bias, and show that both can be replaced by inference-time memory retrieval and permutation aggregation.
First, for memory retrieval, we draw on the framework of \citet{deng2022recent}, which decomposes the disconnect between natural language and SQL structure into \textit{encoding} (understanding natural language semantics), \textit{translating} (mapping those semantics to SQL), and \textit{decoding} (generating executable SQL).
To address this challenge, PAS-SQL~\cite{kong2026bridginggap} extracts question structures and maps each question phrase to database schemas, then uses both to generate SQL.
Second, for optimizing permutation aggregation, we leverage execution feedbacks to group candidates with same results to drastically reduce permutation space from $O(n!)$ to $O(g!)$, where $g$ denotes the number of groups such that $g \ll n$.

Unlike prior work, \ours uses retrieved selection criteria and permutes candidates only within execution-result groups, with confidence-based comparison. 
This design improves both accuracy and efficiency over the corresponding baselines.
\section{Problem Setup}
\label{sec:setup}

\textbf{Input.} We start with a natural language question $x$, database schema $S$, and retrieved memories $M$.
A candidate generator produces a set of SQL candidates $C=\{q_i\}_{i=1}^n$.\footnote{We experiment with $n{=}8$ and $n{=}32$ candidates.}
Each SQL query $q_i$ is executed to produce an execution result $e_i$ (result table, empty result, or error), which is also provided as input to the selector with $q_i$.
Each execution result is computed once and cached for reuse across all sliding windows and permutations.

\noindent \textbf{Output.} Our selector chooses a single SQL query $\hat{q}\in C$ that maximizes correctness (measured by execution match).

We evaluate execution accuracy (exact match) across benchmarks, with BIRD as the primary benchmark~\cite{li2024bird}. We additionally measure efficiency, by the number of selector LLM calls per question (or input tokens per question).
\section{Proposed Method}
\label{sec:method}

In this section, we present \ours, a fine-tuning-free listwise selection framework that replaces training with inference-time strategies. 
Specifically, we use memory retrieval (Section~\ref{sec:memory_aug}) to provide selection criteria and permutation-based aggregation to mitigate positional bias (Section~\ref{sec:positional_bias}).

\begin{figure}[t]
  \centering
  \includegraphics[width=\linewidth]{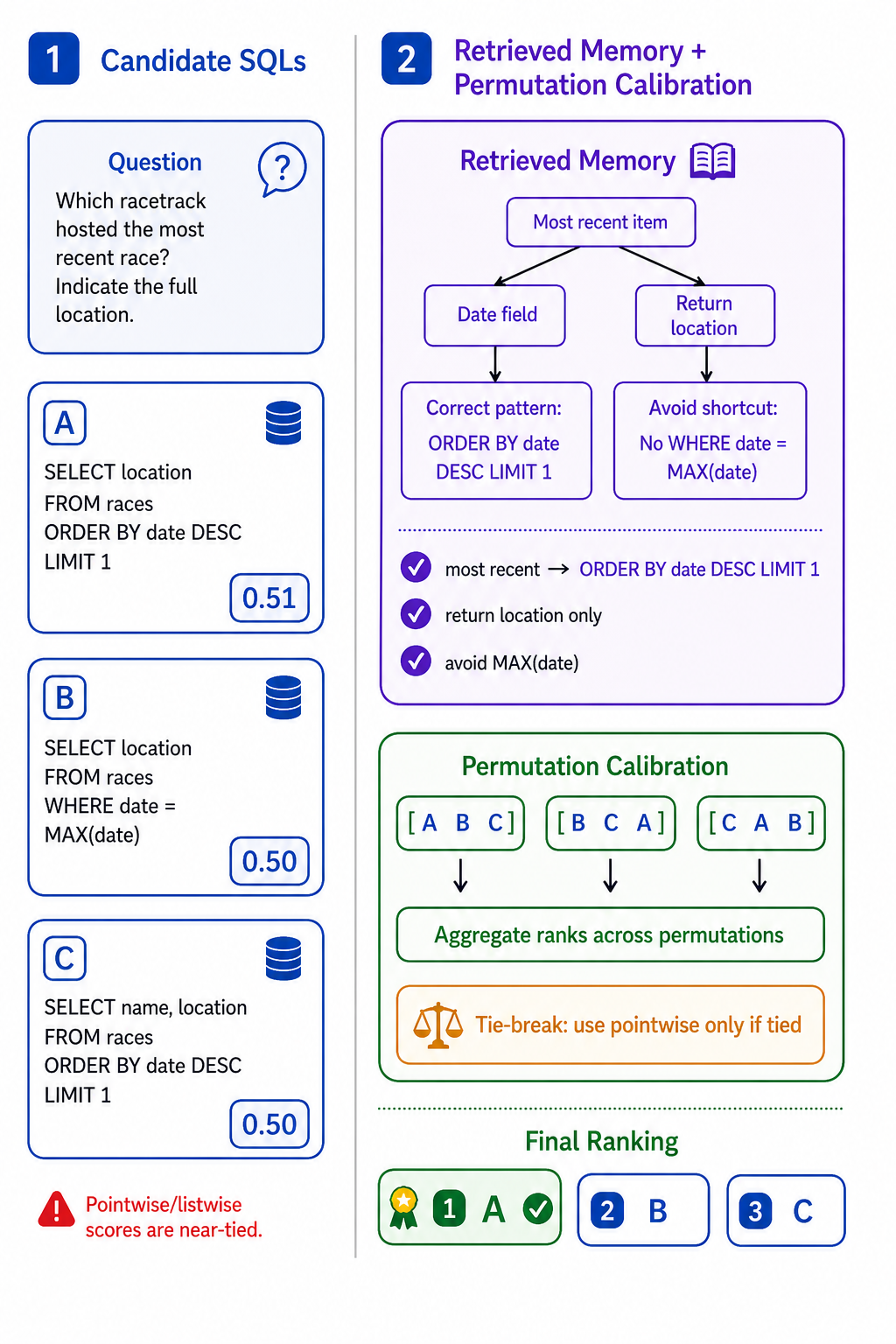}
  \caption{
  Overview of our fine-tuning-free listwise selection framework. Given a question, multiple SQL candidates are generated. To select, the selector retrieves structured memories that provide explicit criteria for validating candidate queries (e.g., schema grounding, ordering patterns, and constraints). To address positional bias, candidates are evaluated across multiple permutations and their rankings are aggregated, with a pointwise tie-break when needed. 
  }
  \label{fig:overview}
\end{figure}

\subsection{Fine-tuning-free Selection with Memory}
\label{sec:memory_aug}

As an alternative to training a selector, we may store the full selection history~\cite{packer2023memgpt, lee2026beyond}, but using histories is difficult under limited context.
Instead, we propose generating a compact memory, avoiding the need for selector fine-tuning.

\paragraph{Step 1: Memory Generation.}
To select the correct candidate, the selector must verify whether the \textit{semantic gap}~\cite{guo2022kggap, kong2026bridginggap} between $x$ and $q_i$ is resolved.
We therefore store how natural language expressions map to SQL operations and schema elements as reusable memories:
\begin{equation}
    M = \{m_j\}_{j=1}^{|\mathcal{D}|}, \quad m_j = f_{\mathrm{mem}}(x_j, S_j, q_j^*),
\end{equation}
where $f_{\mathrm{mem}}$ generates $m_j$ from $(x_j, S_j, q_j^*)$ using the prompt in Figure~\ref{fig:memory_prompt}.\footnote{We use the same LLM as the selector for $f_{\mathrm{mem}}$, to avoid using a separate model that provides no additional gains.}
Borrowing terms from \citet{deng2022recent}, each memory $m_j$ is organized into three groups:
\begin{itemize}
    \item \textbf{Encoding} captures how natural language phrases are grounded to the database schema and conditions.
    \item \textbf{Translating} captures how the grounded meaning is converted into SQL operations.
    \item \textbf{Decoding} captures how the final SQL output should be formed and validated.
\end{itemize}
Figure~\ref{fig:overview} illustrates an example in which the generated candidates are superficially similar, making them difficult for a listwise selector to distinguish. The retrieved memory supplies complementary criteria across the three groups: an encoding criterion for the relevant date field, a translating criterion such as \texttt{ORDER BY ... DESC LIMIT 1} for recency, and a decoding criterion specifying that the location should be returned. Together, these criteria guide the selector toward the correct SQL candidate.
Definitions of the memory keys in each group are provided in Table~\ref{tab:memory_keys}.

\begin{table*}[t]
\centering
\footnotesize
\setlength{\tabcolsep}{4pt}
\renewcommand{\arraystretch}{1.25}
\begin{tabular}{@{}l l p{5.2cm} p{4.8cm}@{}}
\toprule
\textbf{Group} & \textbf{Specification Key} & \textbf{Description} & \textbf{Covered SQL Keywords / Constructs} \\
\midrule

\multirow{3}{*}{Encoding}
  & \texttt{schema\_grounding}
  & Maps question phrases to tables/columns, including confusable ones.
  & (All table and column references.) \\

  & \texttt{join\_path}
  & Specifies required join paths and join types between tables.
  & \texttt{JOIN}, \texttt{INNER JOIN}, \texttt{LEFT JOIN}, \texttt{CROSS JOIN} \\

  & \texttt{filter\_semantics}
  & Identifies WHERE conditions and NULL-related predicates.
  & \texttt{WHERE}, \texttt{IN}, \texttt{LIKE}, \texttt{BETWEEN},
    \texttt{IS NULL}, \texttt{IS NOT NULL} \\

\midrule

\multirow{3}{*}{Translating}
  & \texttt{aggregation}
  & Defines grouping columns, aggregate functions, and HAVING conditions.
  & \texttt{GROUP BY}, \texttt{HAVING}, \texttt{COUNT}, \texttt{SUM},
    \texttt{AVG}, \texttt{MIN}, \texttt{MAX}, \texttt{DISTINCT} \\

  & \texttt{ordering\_and\_scope}
  & Specifies ordering direction and row-count constraints.
  & \texttt{ORDER BY}, \texttt{ASC}, \texttt{DESC}, \texttt{LIMIT}, \texttt{OFFSET} \\

  & \texttt{conditional\_and\_null}
  & Describes conditional branching and null-safe value expressions.
  & \texttt{CASE}, \texttt{IIF}, \texttt{COALESCE}, \texttt{IFNULL}, \texttt{NULLIF} \\

\midrule

\multirow{2}{*}{Decoding}
  & \texttt{output\_form}
  & Specifies result shape, return columns, and value format.
  & \texttt{SELECT}, \texttt{DISTINCT} \\

  & \texttt{query\_constraints}
  & Meta-level constraints to avoid spurious joins or wrong aggregations.
  & (Meta-level; no specific SQL syntax.) \\

\midrule

  & \texttt{extra\_keywords}
  & Covers advanced SQLite constructs beyond the core keys. Also included as \texttt{[]} when empty.
  & All other SQL keywords and constructs not covered by the core keys above. \\

\bottomrule
\end{tabular}
\caption{Specification keys and their roles in the Encoding--Translating--Decoding framework for Text-to-SQL selection.}
\label{tab:memory_keys}
\end{table*}

\paragraph{Step 2: Memory Retrieval.}
Each memory is generated from a question in the training set, so we retrieve memories based on the similarity of the question.
Given a test question $x$, we retrieve the top-$k$ most relevant memories using a dense retriever:
\begin{equation}
    \mathcal{J}^* = \underset{|\mathcal{J}|=k}{\arg\mathrm{top}\text{-}k} \,
    \mathrm{sim}(\mathbf{h}_x, \mathbf{h}_{x_j}),
\end{equation}
where $\mathbf{h}_x$ and $\mathbf{h}_{x_j}$ are dense embeddings of $x$ and $x_j$.
Rather than fixing $k$, we include as many memories as fit within the context limit of the selector to provide diverse perspectives for selection.
The retrieved memories $\mathcal{M}(x) = \{m_j\}_{j \in \mathcal{J}^*}$ are prepended to the prompt.

\paragraph{Step 3: Listwise selection.}
After retrieval, we initialize a candidate order and apply listwise reranking using the retrieved memories.
We set the initial order by placing candidates with more frequent execution results first, as majority voting~\cite{cscsql} is a reliable prior for correctness.
We apply a sliding window of size $w$ and stride $s$ from back to front.\footnote{In our experiments, we use $w=8$ and $s=4$. Thus, each selector call contains at most 8 candidates even when $n=32$.}
For each window $W_t$, the selector produces a local ranking:
\begin{equation}
    \pi_t = f_{\mathrm{list}}(x, S, \mathcal{M}(x), \{(q_i, e_i)\}_{q_i \in W_t}),
\end{equation}
and updates the order by $R[W_t] \leftarrow \pi_t$.
The retrieved memories $\mathcal{M}(x)$ are included once per prompt.
This allows the selector to verify all candidates in the window against the same specifications simultaneously.

\subsection{Fine-tuning-free Bias Mitigation with Permutation}
\label{sec:positional_bias}

To reduce positional bias without selector fine-tuning, we aggregate selection results across multiple candidate orderings.
Prior permutation-based methods~\cite{zeng2024llm, lee2025inference} mitigate positional bias by evaluating candidates across multiple orderings.
Our goal is to mitigate positional bias.
However, they require a large number of such evaluations to be effective,
e.g., up to $O(n!)$ permutations.
Thus, we propose a method that reduces positional bias efficiently with far fewer runs.
Table~\ref{tab:positional_bias} confirms that preserving the between-group ordering leads to better accuracy and less positional bias than prior methods.

We first partition candidates into groups based on execution outcomes, treating each group as a unit during permutation. 
This reduces the permutation space from individual candidates to groups, e.g., $O(g!)$, typically $g \ll n$.\footnote{In our experiments with $n=8$ candidates, the average $g$ was approximately $2$.}
We then consider individual elements only when distinguishing between groups is necessary.

This two-level strategy significantly reduces the number of required permutations while preserving the ability to resolve fine-grained differences between competing candidates.

\paragraph{Group-Based Permutation.}
We propose a group-based permutation strategy to reduce the number of required selection runs.
As in Step 3 of Section~\ref{sec:memory_aug}, we group candidates by their execution results and place larger groups first.
Prior method~\cite{tang2024found} shuffles all candidates globally because they lack a reliable prior over candidate importance, requiring more runs for stable results.
In Text-to-SQL, majority voting serves as a strong prior for correctness. 
Preserving this signal avoids noisy rankings from uninformative orderings.
We therefore fix the ordering between groups and shuffle only within each group.

We rerank candidates by their average rank and apply a confidence estimation procedure to determine whether the top-1 candidate is better than the top-2 candidate.
For each candidate $c_i$, we compute the average rank $\mu_i$ across $K$ runs:
\begin{equation}
    \mu_i = \frac{1}{K}\sum_{k=1}^{K} r_i^{(k)}
\end{equation}
where $r_i^{(k)}$ is the rank of $c_i$ in the k-th run, and candidates are sorted by $\mu_i$ in ascending order.
Let $a$ and $b$ denote the top-1 and top-2 candidates after sorting.
We estimate confidence using pairwise rank differences $d_{a,b}^{(k)} = r_b^{(k)} - r_a^{(k)}$ and compute the confidence score $P(a > b)$ as:\footnote{We use paired differences rather than individual rank variances, since ranks within the same run are not independent. We use this score as a lightweight ranking-confidence heuristic rather than as a formal hypothesis test.}
\begin{equation}
    P(a > b) =
    T_{K-1}\!\left(
    \frac{\bar{d}_{a,b}}{s_{a,b}/\sqrt{K}}
    \right),
\end{equation}
where $\bar{d}_{a,b}$ and $s_{a,b}$ are the mean and standard deviation of $\{d_{a,b}^{(k)}\}$, and $T_{K-1}(\cdot)$ is the CDF of Student's $t$-distribution with $K-1$ degrees of freedom.
A tie is declared when $P(a > b) < \tau$.\footnote{We set $\tau = 0.95$ in all experiments.}

\paragraph{Tie-Breaking with Pointwise Selector.}
When a tie is declared, we can resolve it using a pointwise selector as an optional secondary step.
We score only the tied candidates independently using a pointwise reward model and select the one with the higher score.
Since the pointwise selector is invoked only when a tie occurs, the dominant computational path remains listwise operations.

\section{Experiments}
\label{sec:experiments}
\begin{table*}[t]
\centering
\small
\setlength{\tabcolsep}{3pt}
\renewcommand{\arraystretch}{1.15}
\resizebox{\textwidth}{!}{%
\begin{tabular}{@{}l*{12}{r}@{}}
\toprule
\textbf{Generator}
& \multicolumn{6}{c}{\textbf{Agentar-Scale-SQL-Generation-32B}}
& \multicolumn{6}{c}{\textbf{Arctic-Text2SQL-R1-7B}} \\
\cmidrule(lr){2-7}\cmidrule(lr){8-13}
& \multicolumn{3}{c}{$n{=}8$} & \multicolumn{3}{c}{$n{=}32$}
& \multicolumn{3}{c}{$n{=}8$} & \multicolumn{3}{c}{$n{=}32$} \\
\cmidrule(lr){2-4}\cmidrule(lr){5-7}\cmidrule(lr){8-10}\cmidrule(lr){11-13}
& \textbf{Acc. $\uparrow$} & \textbf{Calls $\downarrow$} & \textbf{Tokens $\downarrow$}
& \textbf{Acc. $\uparrow$} & \textbf{Calls $\downarrow$} & \textbf{Tokens $\downarrow$}
& \textbf{Acc. $\uparrow$} & \textbf{Calls $\downarrow$} & \textbf{Tokens}
& \textbf{Acc. $\uparrow$} & \textbf{Calls $\downarrow$} & \textbf{Tokens $\downarrow$} \\
\midrule

\multicolumn{13}{@{}l}{\textit{\textbf{Baselines}}} \\
~~ greedy & 69.95 & -- & -- & 69.95 & -- & -- & 68.19 & -- & -- & 68.19 & -- & -- \\
~~ majority voting~\cite{cscsql} & 71.38 & -- & -- & 70.93 & -- & -- & 70.08 & -- & -- & 69.95 & -- & -- \\
\midrule

\multicolumn{13}{@{}l}{\textit{\textbf{Single Selector}}} \\
~~ pointwise~\citep{contextual2025} & 71.45 & 8.00 & 16{,}793 & 71.71 & 32.00 & 67{,}299 & 70.27 & 8.00 & 16{,}798 & 70.47 & 32.00 & 67{,}168 \\
~~ pairwise~\cite{pourreza2025chase} & 71.25 & 6.11 & 14{,}251 & 71.06 & 129.93 & 309{,}564 & 69.30 & 9.71 & 23{,}498 & 68.12 & 184.59 & 443{,}713 \\
\rowcolor{gray!15}
~~ listwise w/ Memory (ours) & 72.23 & \textbf{0.71} & \textbf{3{,}347} & 72.75 & \textbf{4.72} & \textbf{22{,}190} & 71.90 & \textbf{1.12} & \textbf{5{,}147} & 72.10 & \textbf{5.91} & \textbf{27{,}440} \\
\midrule

\multicolumn{13}{@{}l}{\textit{\textbf{Multiple Selectors}}} \\
~~ \rthree (not trained)~\cite{r3-sql} & 71.51 & 14.11 & 31{,}044 & 71.97 & 161.93 & 376{,}863 & 69.36 & 17.71 & 40{,}295 & 69.56 & 216.59 & 510,881 \\
\rowcolor{gray!15}
~~ \ours (ours) & \textbf{72.62} & 2.97 & 15{,}765 & \textbf{73.08} & 19.04 & 98{,}234 & \textbf{72.16} & 4.65 & 23{,}567 & \textbf{72.62} & 23.85 & 122,985 \\
\bottomrule
\end{tabular}
}
\caption{Full results across generation and selection strategies on BIRD-dev. Acc.\ = execution accuracy (\%). Calls = \# calls. Tokens = \# tokens. `--' = not reported or not applicable.}
\label{tab:generator_results}
\end{table*}
\begin{table*}[t]
\centering
\small
\setlength{\tabcolsep}{3pt}
\renewcommand{\arraystretch}{1.15}
\resizebox{\textwidth}{!}{%
\begin{tabular}{@{}l*{12}{r}@{}}
\toprule
\textbf{Dataset}
& \multicolumn{6}{c}{\textbf{Spider-test}}
& \multicolumn{6}{c}{\textbf{EHRSQL}} \\
\cmidrule(lr){2-7}\cmidrule(lr){8-13}
& \multicolumn{3}{c}{$n{=}8$} & \multicolumn{3}{c}{$n{=}32$}
& \multicolumn{3}{c}{$n{=}8$} & \multicolumn{3}{c}{$n{=}32$} \\
\cmidrule(lr){2-4}\cmidrule(lr){5-7}\cmidrule(lr){8-10}\cmidrule(lr){11-13}
& \textbf{Acc. $\uparrow$} & \textbf{Calls $\downarrow$} & \textbf{Tokens $\downarrow$}
& \textbf{Acc. $\uparrow$} & \textbf{Calls $\downarrow$} & \textbf{Tokens $\downarrow$}
& \textbf{Acc. $\uparrow$} & \textbf{Calls $\downarrow$} & \textbf{Tokens $\downarrow$}
& \textbf{Acc. $\uparrow$} & \textbf{Calls $\downarrow$} & \textbf{Tokens $\downarrow$} \\
\midrule

\multicolumn{13}{@{}l}{\textit{\textbf{Baselines}}} \\
~~ greedy & 86.40 & -- & -- & 86.40 & -- & -- &36.52 & -- & -- & 36.52 & -- & -- \\
~~ majority voting~\cite{cscsql} & 87.07 & -- & -- & 87.02 & -- & -- & 38.10 & -- & -- & 39.59 & -- & --\\
\midrule

\multicolumn{13}{@{}l}{\textit{\textbf{Single Selector}}} \\
~~ pointwise~\citep{contextual2025} & 86.88 & 8.00 & 4,848 & 86.72 & 32.00 & 19,399 & 38.79 & 8.00 & 24,611 & 43.52 & 32.00 & 98,579 \\
~~ pairwise~\cite{pourreza2025chase} & 86.83 & 4.54 & 4,191 & 86.63 & 74.40 & 69,682 & 38.23 & 22.61 & 74,708 & 43.67 & 408.11 & 1,341,930\\
\rowcolor{gray!15}
~~ listwise w/ Memory (ours) & 87.36 & \textbf{0.57} & \textbf{1,377}& 87.40 & \textbf{3.13} & \textbf{7,612} & 39.08 & \textbf{2.11} & \textbf{10,340} & 44.20 & \textbf{9.84} & \textbf{48,060} \\
\midrule

\multicolumn{13}{@{}l}{\textit{\textbf{Multiple Selectors}}} \\
~~ \rthree (not trained)~\cite{r3-sql} & 87.21 & 12.54 & 9,040 & 86.82 & 106.40 & 89,081 & 39.25 & 30.61 & 99,319 & 44.03 & 440.11 & 1,440,510 \\
\rowcolor{gray!15}
~~ \ours (ours) & \textbf{87.49} & 2.38 & 6,013 & \textbf{87.59} & 12.63 & 32,659 & \textbf{39.93} & 8.77 & 49,516 & \textbf{44.71} & 39.70 & 228,070 \\
\bottomrule
\end{tabular}
}
\caption{Results on Spider-test and EHRSQL using Arctic-Text2SQL-R1-7B as the generator. Acc.\ = execution accuracy (\%). Calls = \# calls. Tokens = \# tokens. `--' = not reported or not applicable.}
\label{tab:other_benchmark_results}
\end{table*}
\subsection{Experimental Setup}
\paragraph{Benchmarks.}
We evaluate on three Text-to-SQL benchmarks: BIRD-dev~\citep{li2024bird}, Spider-test~\citep{yu2018spider}, and EHRSQL~\citep{lee2022ehrsql}.
BIRD is our primary benchmark, consisting of 1,534 development queries over large-scale, realistic databases.
Spider-test is a widely used cross-domain Text-to-SQL benchmark consisting of 2,147 queries, while EHRSQL contains 1,008 questions in the electronic health record domain.

\paragraph{Metrics.}
We report execution accuracy (Acc.), which measures whether the predicted SQL produces the same result as the gold SQL.
We also report the average number of LLM calls (Calls) and input tokens per query (Tokens) to assess computational efficiency.

\paragraph{Baselines.}
We compare against three selection baselines: pointwise, pairwise, and a multiple-selector approach using both (e.g. \rthree and \ours).
For the \textbf{pointwise} selector, we use Contextual-RM-32B~\citep{contextual2025}\footnote{\href{https://huggingface.co/ContextualAI/ctx-bird-reward-250121}{Contextual-RM-32B}}, a strong reward model trained from Qwen2.5-Coder-32B-Instruct and released by Contextual-SQL.
For the \textbf{pairwise} selector, we follow the approach of CHASE-SQL~\cite{pourreza2025chase}.
In this method, candidates with identical execution results are grouped and not compared against each other, and only cross-group pairs are compared, by using the prompt in Figure~\ref{fig:pairwise_prompt_template}.
The final ranking of pairwise is obtained by aggregating these pairwise outcomes.
For \textbf{\rthree}~\citep{r3-sql}, we reproduce its selection algorithm by using the same generator and selector as ours, since neither CHASE-SQL nor \rthree has released their original models.
This follows the same evaluation protocol as \rthree, which also uses a shared selector to isolate the contribution of the selection algorithm for fair comparison.
For all listwise and pairwise components, we use Qwen3-Coder-30B-A3B-Instruct\footnote{\href{https://huggingface.co/Qwen/Qwen3-Coder-30B-A3B-Instruct}{Qwen3-Coder-30B-A3B-Instruct}} as the selector.
For \ours, Contextual-RM-32B is used only for optional tie-breaking. Appendix~\ref{sec:additional_results} reports results without tie-breaking and with Qwen3-Coder-30B-A3B-Instruct as the tie-breaker.

\paragraph{Generators.}
We use two SQL generators to assess robustness across different candidate pools: Agentar-Scale-SQL-Generation-32B~\citep{wang2025agentar} and Arctic-Text2SQL-R1-7B~\citep{arctic2025text2sql}.
For each question, we generate either $n{=}8$ or $n{=}32$ candidates and evaluate selection performance on top of these fixed candidate pools.

\subsection{Implementation Details}
\paragraph{Memory Retrieval.}
We generate memories from training questions following the procedure described in Section~\ref{sec:memory_aug}.
At inference time, we retrieve the top-$k$ most relevant memories using bge-m3~\cite{chen2024bge} as the dense retriever, based on question similarity.
Rather than fixing $k$, we include as many memories as fit within the selector's context limit. For BIRD and Spider, we construct the memory bank from each benchmark's training split.Because EHRSQL does not provide a training split, we use a memory bank constructed from the combined BIRD and Spider training splits.

\paragraph{Selection Prompts.}
The listwise selector uses the prompt template shown in Figure~\ref{fig:listwise_prompt_template}.

\paragraph{Experimental Infrastructure.}
All experiments are conducted on a single node equipped with 1 NVIDIA RTX PRO 6000 GPU.

\subsection{Experimental Results}

Tables~\ref{tab:generator_results} and~\ref{tab:other_benchmark_results} present the results across all benchmarks.
Our method consistently outperforms all baselines in accuracy while requiring substantially fewer LLM calls and input tokens.

\paragraph{Accuracy.}
Our full method achieves the highest execution accuracy across all settings.
On BIRD-dev with Agentar-32B and $n{=}32$, it reaches 73.08\%, outperforming \rthree by 1.11 points and pairwise by 2.02 points.
The gains are consistent across both generators and both candidate pool sizes.
Notably, our single listwise selector already surpasses \rthree in most settings, despite \rthree combining multiple selectors.

\paragraph{Efficiency.}
Our listwise selector reduces computational cost significantly compared to pairwise and \rthree.
On BIRD-dev with $n{=}32$, pairwise requires on average 184.59 calls and 443,713 tokens per query with Arctic-R1-7B, whereas our listwise selector uses only 5.91 calls and 27,440 tokens.
Compared to \rthree, our full method uses 9.07$\times$ fewer calls and 4.16$\times$ fewer tokens in the same setting.
On EHRSQL with $n{=}32$, the token reduction is even more pronounced: pairwise consumes over 1.3M tokens per query, while our listwise selector requires only 48,060.

\paragraph{Generalization.}
On Spider-test and EHRSQL, our method also outperforms all baselines.
With $n{=}32$ on Spider-test, our full method achieves 87.59\% accuracy, surpassing \rthree by 0.77 points using 8.43$\times$ fewer calls.
On EHRSQL with $n{=}32$, our method reaches 44.71\%, improving over \rthree by 0.68 points while requiring 11.09$\times$ fewer calls.
These results confirm that the advantage of listwise selection holds beyond the primary benchmark.

\section{Analysis}
\label{sec:analysis}

\subsection{Ablation Study}
\begin{table}[t]
    \centering
    \small
    \setlength{\tabcolsep}{6pt}
    \renewcommand{\arraystretch}{1.12}
    \begin{tabular}{lcc}
        \toprule
        \textbf{Method} & \textbf{Agentar} & \textbf{Arctic} \\
        \midrule
        \ours & \textbf{72.62} & \textbf{72.16} \\
        \phantom{0}\textit{w/o Permutation} & 72.23 & 71.90 \\
        \phantom{0}\textit{w/o Memory} & 71.94 & 71.74 \\
        \phantom{00}\textit{w/o Memory \& Permutation} & 71.64 & 71.32 \\
        \bottomrule
    \end{tabular}
    \caption{Ablation study on BIRD-dev with $n{=}8$ candidates generated by Agentar-Scale-SQL-Generation-32B (Agentar) and Arctic-Text2SQL-R1-7B (Arctic). We report execution accuracy (\%).}
    \label{tab:ablation}
\end{table}

To verify the contribution of each component in our method, we conduct an ablation study on BIRD-dev with $n=8$ candidates.
Our full system consists of two components: memory retrieval, which provides structured selection criteria, and permutation-based aggregation, which mitigates positional bias.
As shown in Table~\ref{tab:ablation}, the full \ours system achieves the best performance on both generators, with 72.62 on Agentar-Scale-SQL-Generation-32B~\cite{wang2025agentar} (Agentar) and 72.16 on Arctic-Text2SQL-R1-7B (Arctic)~\cite{arctic2025text2sql}.
Removing either component consistently lowers accuracy, which shows that both components contribute to the final selection performance.

The two components provide complementary benefits.
Without Permutation, the average accuracy decreases from 72.39 to 72.07, resulting in a drop of 0.32 points. 
Without Memory, the average accuracy decreases to 71.84, resulting in a larger drop of 0.55 points.

\subsection{Memories with a Strong Code Selector}

One concern is that a strong agentic code model may already form memory-like criteria during its own reasoning.
Table~\ref{tab:rubric_gain} shows that explicit retrieved memories still help.
With a \texttt{gpt-5.1-codex-mini} selector, accuracy rises from 71.90\% to 72.23\% for Agentar-32B and from 70.01\% to 70.40\% for Arctic-R1-7B.

The gain is modest, but it is consistent across both generators.
This suggests that the memory complements the selector's internal reasoning rather than duplicating it.
It is closer to a retrieved evaluation template than to open-ended memory.
The selector does not revisit past trajectories or maintain long interaction histories.
Instead, it receives a short structured criterion set once and applies it to the current candidate list.

\begin{table}[t]
\centering
\small
\setlength{\tabcolsep}{6pt}
\begin{tabular}{@{}lcc@{}}
\toprule
\textbf{Generator} & \textbf{Listwise} & \textbf{+ Memory} \\
\midrule
Agentar-32B & 71.90 & \textbf{72.23} (+0.33) \\
Arctic-R1-7B & 70.01 & \textbf{70.40} (+0.39) \\
\bottomrule
\end{tabular}
\caption{Memory gains on BIRD-dev with a \texttt{gpt-5.1-codex-mini} selector and $n{=}8$ candidates per query.}
\label{tab:rubric_gain}
\end{table}

\subsection{Mitigating Positional Bias}

We examine whether our method selects the same candidate regardless of the order in which candidates are presented. 
To this end, we provide the model with two candidate sets that contain the same candidates but in different orderings and measure how often the model selects the same candidate from both. 
We refer to this agreement rate as \textit{consistency} and separately measure whether the agreed-upon selection is actually correct, which we call \textit{consistency \& correct}. 
We set the candidate set size to $n=8$.
As a reference, random selection yields a consistency of 12.58\%.
Each compared method applies permutations according to its own strategy before aggregating the final selection.

Table~\ref{tab:positional_bias} demonstrates that each component of our aggregation method contributes to mitigating positional bias. The baseline listwise method achieves a consistency of 18.98\%, only modestly above the random baseline. Adding global permutation improves this to 23.04\%, but the gain remains limited. Group-based permutation, by contrast, raises consistency substantially to 29.04\%, a gain of 6.0 percentage points over global permutation. The tie-break strategy further pushes consistency to 33.17\% with a corresponding consistency \& correct of 22.63\%. These results show that both components contribute meaningfully and that group-based permutation accounts for the larger share of the improvement.

\begin{figure}[t]
\centering
\includegraphics[width=\columnwidth]{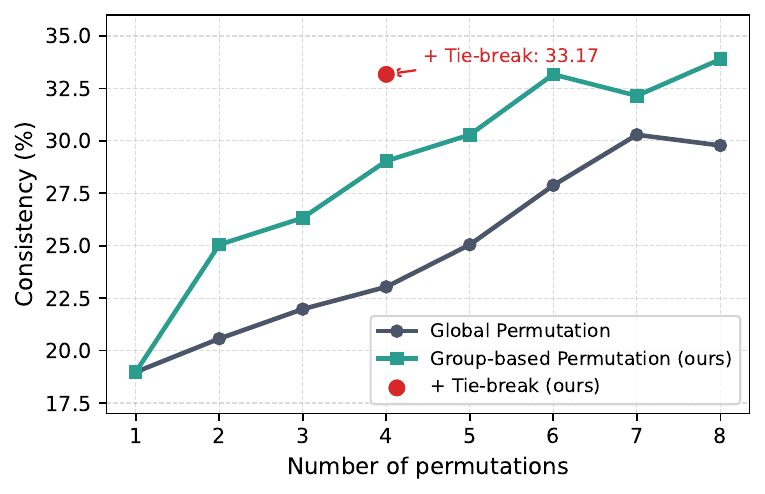}
\caption{Consistency under different numbers of permutations. Group-wise permutation improves selection consistency over global permutation, and the tie-break strategy further improves consistency at four permutations.}
\label{fig:permutation_consistency}
\end{figure}

Figure~\ref{fig:permutation_consistency} further reveals how the two permutation strategies behave as the number of permutations increases. Global permutation yields lower consistency throughout and improves only slowly with more permutations. Group-based permutation achieves higher consistency from the start and saturates more quickly. This faster saturation is because group-based permutation avoids noisy orderings that can mislead the model. The tie-break strategy makes this even more efficient. At four permutations, applying the tie-break reaches 33.17\%, which is comparable to running group-based permutation eight times without it. The tie-break thus achieves the benefit of doubling the number of permutations without requiring additional inference.

\begin{table}[t]
\centering
\small
\setlength{\tabcolsep}{4pt}
\resizebox{\columnwidth}{!}{%
\begin{tabular}{@{}lcc@{}}
\toprule
\textbf{Method} &
\shortstack{\textbf{Consistency} \\ \textbf{(\%) $\uparrow$}} &
\shortstack{\textbf{Consistency \&} \\ \textbf{Correct (\%) $\uparrow$}} \\
\midrule
Random & 12.58 & 8.67 \\
\midrule
Listwise & 18.98 & 11.46 \\
\quad + Global Permutation & 23.04 & 14.18 \\
\quad + Group-based Permutation (ours) & 29.04 & 19.18 \\
\quad \quad + Tie-break (ours) & \textbf{33.17} & \textbf{22.63} \\
\bottomrule
\end{tabular}%
}
\caption{Positional-bias analysis with shuffled initial candidate orders ($n=8$). Group-based permutation and tie-break strategies improve selection consistency while mitigating positional bias.}
\label{tab:positional_bias}
\end{table}

\subsection{Comparison with a Fine-Tuned Selector}
\begin{table}[t]
    \centering
    \small
    \setlength{\tabcolsep}{6pt}
    \renewcommand{\arraystretch}{1.12}
    \begin{tabular}{lc}
        \toprule
        \textbf{Method} & \textbf{Acc.} \\
        \midrule
        majority voting & 68.45 \\
        R$^3$-SQL (fine-tuned) & 71.84 \\
        \ours & \textbf{71.90} \\
        \bottomrule
    \end{tabular}
    \caption{Comparison with a fine-tuned selector on BIRD-dev using $n{=}32$ candidates generated by OmniSQL-7B. We report execution accuracy (\%).}
    \label{tab:trained_selector_comparison}
\end{table}

To assess whether fine-tuning-free selection can match fine-tuned selectors, we compare \ours against the reported result of R$^3$-SQL~\cite{r3-sql} using its fine-tuned selector. 
Since the fine-tuned selector of R$^3$-SQL is not publicly released, we rely on its reported result: 71.84\% execution accuracy on BIRD-dev with OmniSQL-7B~\cite{li2025omnisql} as the generator and $n{=}32$ candidates.

As shown in Table~\ref{tab:trained_selector_comparison}, \ours achieves 71.90\% with OmniSQL-7B, slightly surpassing this figure without selector fine-tuning.
This result suggests that our inference-time strategies are sufficient to reach a comparable level of selection quality to a trained selector.
\section{Conclusion}
\label{sec:conclusion}

We present \ours, and (1)~show that listwise selection outperforms pairwise and pointwise baselines for SQL candidate selection.
The selector contrasts SQL structures, execution outcomes, and schema usage across multiple candidates in a single call.
We also (2)~show that retrieved memories further improve listwise selection.
With a strong code-focused selector, the gain remains consistent across multiple generators.
(3)~Listwise selection reduces call complexity from $O(N^2)$ to $O(N)$--$O(N \log N)$, using up to 27.92$\times$ fewer input tokens per query than pairwise selection in our experiments.
\section*{Limitations}
We believe there is still room for improvement on applying advanced coordination techniques for listwise selection, other than just using sliding windows, such as TourRank~\cite{chen2025tourrank}, or Set-based ranking~\cite{zhuang_setwise}. 
We leave the exploration of more advanced coordination strategies as future work.

The overall execution accuracy (70--72\%) lags behind SOTA systems using proprietary models (75--76\%), partly due to generator quality rather than selection strategy.
Our results characterize selector performance on the same candidate pools rather than end-to-end SOTA performance on BIRD.

For enterprise-scale databases with hundreds of tables, we assume upstream schema linking to prune the schema before generation and selection~\citep{talaei2024chess}, while schema-scale context management itself is beyond the scope of our selector.

We focus on Text-to-SQL because it is a challenging listwise selection setting involving long schemas, execution results, and multiple candidate SQL queries. Exploring whether our memory-guided selection and group-based permutation generalize beyond Text-to-SQL is left for future work.

\section*{Acknowledgements}
This work was supported by 
the National Research Foundation of Korea (NRF) grant funded by the Korea government (MSIT) (No. RS-2024-00414981),
the MSIT (Ministry of Science and ICT), Korea, under the ITRC (Information Technology Research Center) support program (IITP-2025-2020-0-01789) supervised by the IITP (Institute for Information \& Communications Technology Planning \& Evaluation), and
Institute of Information \& communications Technology Planning \& Evaluation (IITP) grant funded by the Korea government(MSIT) [NO.RS-2021-II211343, Artificial Intelligence Graduate School Program (Seoul National University)].

\bibliography{custom}

@inproceedings{pourreza2025chase,
title={{CHASE}-{SQL}: Multi-Path Reasoning and Preference Optimized Candidate Selection in Text-to-{SQL}},
author={Mohammadreza Pourreza and Hailong Li and Ruoxi Sun and Yeounoh Chung and Shayan Talaei and Gaurav Tarlok Kakkar and Yu Gan and Amin Saberi and Fatma Ozcan and Sercan O Arik},
booktitle={The Thirteenth International Conference on Learning Representations},
year={2025},
url={https://openreview.net/forum?id=CvGqMD5OtX}
}

@article{li2024bird,
  title={Can {LLM} Already Serve as A Database Interface? {A} {BIg} Bench for Large-Scale Database Grounded Text-to-{SQL}s},
  author={Li, Jinyang and Hui, Binyuan and Qu, Ge and Yang, Jiaxi and Li, Binhua and Li, Bowen and Wang, Bailin and Qin, Bowen and Geng, Ruiying and Huo, Nan and others},
  journal={Advances in Neural Information Processing Systems},
  volume={36},
  year={2024}
}

@inproceedings{lee2024mcs,
  title={{MCS-SQL}: Leveraging Multiple Prompts and Multiple-Choice Selection for Text-to-{SQL} Generation},
  author={Lee, Dongjun and Park, Choongwon and Kim, Jaehyuk and Park, Heesoo},
  booktitle={Proceedings of the 31st International Conference on Computational Linguistics},
  pages={337--353},
  address={Abu Dhabi, UAE},
  publisher={Association for Computational Linguistics},
  month={January},
  year={2025}
}

@article{talaei2024chess,
  title={{CHESS}: Contextual Harnessing for Efficient {SQL} Synthesis},
  author={Talaei, Shayan and Pourreza, Mohammadreza and Chang, Yu-Chen and Mirhoseini, Azalia and Saberi, Amin},
  journal={arXiv preprint arXiv:2405.16755},
  year={2024}
}

@inproceedings{yu2018spider,
  title={Spider: A large-scale human-labeled dataset for complex and cross-domain semantic parsing and text-to-sql task},
  author={Yu, Tao and Zhang, Rui and Yang, Kai and Yasunaga, Michihiro and Wang, Dongxu and Li, Zifan and Ma, James and Li, Irene and Yao, Qingning and Roman, Shanelle and others},
  booktitle={Proceedings of the 2018 conference on empirical methods in natural language processing},
  pages={3911--3921},
  year={2018}
}

@inproceedings{arctic2025text2sql,
  title={Arctic-text2sql-r1: Simple rewards, strong reasoning in text-to-sql},
  author={Yao, Zhewei and Sun, Guoheng and Borchmann, {\L}ukasz and Shen, Zheyu and Deng, Minghang and Zhai, Bohan and Zhang, Hao and Li, Ang and He, Yuxiong},
  booktitle={Findings of the Association for Computational Linguistics: ACL 2026},
  pages={26966--26995},
  year={2026}
}

@inproceedings{zhuang_setwise,
  title={A Setwise Approach for Effective and Highly Efficient Zero-shot Ranking with Large Language Models},
  author={Zhuang, Shengyao and Zhuang, Honglei and Koopman, Bevan and Zuccon, Guido},
  booktitle={Proceedings of the 47th International ACM SIGIR Conference on Research and Development in Information Retrieval},
  pages={38--47},
  year={2024},
  doi={10.1145/3626772.3657813}
}

@misc{genasql,
      title={Cheaper, Better, Faster, Stronger: Robust Text-to-SQL without Chain-of-Thought or Fine-Tuning}, 
      author={Yusuf Denizay Dönder and Derek Hommel and Andrea W Wen-Yi and David Mimno and Unso Eun Seo Jo},
      year={2025},
      eprint={2505.14174},
      archivePrefix={arXiv},
      primaryClass={cs.CL},
      url={https://arxiv.org/abs/2505.14174}, 
}

@article{xiyansql,
   title={XiYan-SQL: A Novel Multi-Generator Framework for Text-to-SQL},
   volume={38},
   ISSN={2326-3865},
   url={http://dx.doi.org/10.1109/TKDE.2026.3657851},
   DOI={10.1109/tkde.2026.3657851},
   number={4},
   journal={IEEE Transactions on Knowledge and Data Engineering},
   publisher={Institute of Electrical and Electronics Engineers (IEEE)},
   author={Liu, Yifu and Zhu, Yin and Gao, Yingqi and Luo, Zhiling and Li, Xiaoxia and Shi, Xiaorong and Hong, Yuntao and Gao, Jinyang and Li, Yu and Ding, Bolin and Zhou, Jingren},
   year={2026},
   month=apr, pages={2474–2487} }

@inproceedings{cscsql,
  title={Csc-sql: Corrective self-consistency in text-to-sql via reinforcement learning},
  author={Sheng, Lei and Shuai, Xu Shuai},
  booktitle={Proceedings of the 14th International Joint Conference on Natural Language Processing and the 4th Conference of the Asia-Pacific Chapter of the Association for Computational Linguistics},
  pages={1473--1496},
  year={2025}
}

@inproceedings{r3-sql,
  title={R3-SQL: Ranking Reward and Resampling for Text-to-SQL},
  author={Han, Hojae and Jeong, Yeonseok and Hwang, Seung-won and Yao, Zhewei and He, Yuxiong},
  booktitle={Findings of the Association for Computational Linguistics: ACL 2026},
  pages={43257--43275},
  year={2026}
}

@article{liu2024lost,
  title={Lost in the middle: How language models use long contexts},
  author={Liu, Nelson F and Lin, Kevin and Hewitt, John and Paranjape, Ashwin and Bevilacqua, Michele and Petroni, Fabio and Liang, Percy},
  journal={Transactions of the association for computational linguistics},
  volume={12},
  pages={157--173},
  year={2024}
}

@inproceedings{deng2022recent,
  title={Recent advances in text-to-SQL: A survey of what we have and what we expect},
  author={Deng, Naihao and Chen, Yulong and Zhang, Yue},
  booktitle={Proceedings of the 29th International conference on computational linguistics},
  pages={2166--2187},
  year={2022}
}

@misc{contextual2025,
  author       = {Sheshansh Agrawal and Thien Nguyen},
  title        = {Open-Sourcing the Best Local Text-to-SQL System},
  year         = {2025},
  url          = {https://contextual.ai/blog/open-sourcing-the-best-local-text-to-sql-system/}
}

@article{wang2025agentar,
  title={Agentar-Scale-SQL: Advancing Text-to-SQL through Orchestrated Test-Time Scaling},
  author={Wang, Pengfei and Sun, Baolin and Dong, Xuemei and Dai, Yaxun and Yuan, Hongwei and Chu, Mengdie and Gao, Yingqi and Qi, Xiang and Zhang, Peng and Yan, Ying},
  journal={arXiv preprint arXiv:2509.24403},
  year={2025}
}

@inproceedings{tang2024found,
  title={Found in the middle: Permutation self-consistency improves listwise ranking in large language models},
  author={Tang, Raphael and Zhang, Crystina and Ma, Xueguang and Lin, Jimmy and T{\"u}re, Ferhan},
  booktitle={Proceedings of the 2024 Conference of the North American Chapter of the Association for Computational Linguistics: Human Language Technologies (Volume 1: Long Papers)},
  pages={2327--2340},
  year={2024}
}

@article{lee2022ehrsql,
  title={Ehrsql: A practical text-to-sql benchmark for electronic health records},
  author={Lee, Gyubok and Hwang, Hyeonji and Bae, Seongsu and Kwon, Yeonsu and Shin, Woncheol and Yang, Seongjun and Seo, Minjoon and Kim, Jong-Yeup and Choi, Edward},
  journal={Advances in Neural Information Processing Systems},
  volume={35},
  pages={15589--15601},
  year={2022}
}

@inproceedings{rankgpt,
  title={Is ChatGPT good at search? investigating large language models as re-ranking agents},
  author={Sun, Weiwei and Yan, Lingyong and Ma, Xinyu and Wang, Shuaiqiang and Ren, Pengjie and Chen, Zhumin and Yin, Dawei and Ren, Zhaochun},
  booktitle={Proceedings of the 2023 conference on empirical methods in natural language processing},
  pages={14918--14937},
  year={2023}
}

@article{li2025omnisql,
  title={Omnisql: Synthesizing high-quality text-to-sql data at scale},
  author={Li, Haoyang and Wu, Shang and Zhang, Xiaokang and Huang, Xinmei and Zhang, Jing and Jiang, Fuxin and Wang, Shuai and Zhang, Tieying and Chen, Jianjun and Shi, Rui and others},
  journal={arXiv preprint arXiv:2503.02240},
  year={2025}
}

@article{bai2025judgesql,
  title={JudgeSQL: Reasoning over SQL Candidates with Weighted Consensus Tournament},
  author={Bai, Jiayuan and Pan, Xuan-guang and Tao, Chongyang and Ma, Shuai},
  journal={arXiv preprint arXiv:2510.15560},
  year={2025}
}

@article{tritto2025gradesql,
  title={GradeSQL: Outcome reward models for intelligent Text-to-SQL generation from LLMs},
  author={Tritto, Mattia and Farano, Giuseppe and Di Palma, Dario and Rossiello, Gaetano and Subramanian, Dharmashankar and Narducci, Fedelucio and Di Noia, Tommaso},
  journal={Journal of Intelligent Information Systems},
  pages={1--24},
  year={2026},
  publisher={Springer}
}

@article{packer2023memgpt,
  title={MemGPT: Towards LLMs as Operating Systems},
  author={Packer, Charles and Wooders, Sarah and Lin, Kevin and Fang, Vivian and Patil, Shishir G and Stoica, Ion and Gonzalez, Joseph E},
  journal={arXiv preprint arXiv:2310.08560},
  year={2023}
}

@inproceedings{kong2026bridginggap,
  title={Bridging the gap: transforming natural language questions into sql queries via abstract query pattern and contextual schema markup},
  author={Kong, Yonghui and Hu, Hongbing and Zhang, Dan and Xu, Zhaohui and Wang, Wei},
  booktitle={ICASSP 2026-2026 IEEE International Conference on Acoustics, Speech and Signal Processing (ICASSP)},
  pages={16902--16906},
  year={2026},
  organization={IEEE}
}

@inproceedings{guo2022kggap,
  title={KG-SQL: Hybrid knowledge-guided semantic understanding for text-to-SQL},
  author={Guo, Mengfei and Chen, Yufeng and Xu, Jinan and Zhang, Yujie},
  booktitle={2022 4th International Conference on Natural Language Processing (ICNLP)},
  pages={409--413},
  year={2022},
  organization={IEEE}
}

@inproceedings{long2025precise,
  title={Precise Zero-Shot Pointwise Ranking with LLMs through Post-Aggregated Global Context Information},
  author={Long, Kehan and Li, Shasha and Xu, Chen and Tang, Jintao and Wang, Ting},
  booktitle={Proceedings of the 48th International ACM SIGIR Conference on Research and Development in Information Retrieval},
  pages={2384--2394},
  year={2025}
}

@article{
zeng2024llm,
title={{LLM}-RankFusion:  Mitigating Intrinsic Inconsistency in {LLM}-based Ranking},
author={Yifan Zeng and Ojas Tendolkar and Raymond Baartmans and Qingyun Wu and Lizhong Chen and Huazheng Wang},
journal={Transactions on Machine Learning Research},
issn={2835-8856},
year={2026},
url={https://openreview.net/forum?id=VUY0j74Yes},
note={}
}

@inproceedings{lee2025inference,
  title={Inference scaling for bridging retrieval and augmented generation},
  author={Lee, Youngwon and Hwang, Seung-won and Campos, Daniel F and Gralinski, Filip and Yao, Zhewei and He, Yuxiong},
  booktitle={Findings of the Association for Computational Linguistics: NAACL 2025},
  pages={7324--7339},
  year={2025}
}

@inproceedings{chen2024bge,
    title = "{M}3-Embedding: Multi-Linguality, Multi-Functionality, Multi-Granularity Text Embeddings Through Self-Knowledge Distillation",
    author = "Chen, Jianlyu  and
      Xiao, Shitao  and
      Zhang, Peitian  and
      Luo, Kun  and
      Lian, Defu  and
      Liu, Zheng",
    editor = "Ku, Lun-Wei  and
      Martins, Andre  and
      Srikumar, Vivek",
    booktitle = "Findings of the Association for Computational Linguistics: ACL 2024",
    month = aug,
    year = "2024",
    address = "Bangkok, Thailand",
    publisher = "Association for Computational Linguistics",
    url = "https://aclanthology.org/2024.findings-acl.137",
    doi = "10.18653/v1/2024.findings-acl.137",
    pages = "2318--2335",
}

@inproceedings{chang2023drspider,
  title={{Dr.Spider}: A Diagnostic Evaluation Benchmark towards Text-to-{SQL} Robustness},
  author={Chang, Shuaichen and Wang, Jun and Dong, Mingwen and Pan, Lin and Zhu, Henghui and Li, Alexander Hanbo and Lan, Wuwei and Zhang, Sheng and Jiang, Jiarong and Lilien, Joseph and Ash, Steve and Wang, William Yang and Wang, Zhiguo and Castelli, Vittorio and Ng, Patrick and Xiang, Bing},
  booktitle={The Eleventh International Conference on Learning Representations},
  year={2023}
}

@inproceedings{chen2025tourrank,
  title={TourRank: Utilizing Large Language Models for Documents Ranking with a Tournament-Inspired Strategy},
  author={Chen, Yiqun and Liu, Qi and Zhang, Yi and Sun, Weiwei and Ma, Xinyu and Yang, Wei and Shi, Daiting and Mao, Jiaxin and Yin, Dawei},
  booktitle={Proceedings of the ACM Web Conference 2025},
  year={2025}
}

@inproceedings{lee2026beyond,
  title={Beyond Markovian Forgetfulness: Episodic Memory for Reasoning-Intensive Retrieval},
  author={Lee, Dohyeon and Jeong, Yeonseok and Hwang, Seung-Won},
  booktitle={Proceedings of the 64th Annual Meeting of the Association for Computational Linguistics (Volume 1: Long Papers)},
  pages={37266--37280},
  year={2026}
}

\appendix
\clearpage
\section*{\centering Appendices}

\section{Additional Experiments}
\label{sec:additional_results}

We provide additional experiments to validate \ours from several perspectives.
We first examine whether \ours depends on a separate pointwise reward model for tie-breaking (Section~\ref{sec:tiebreak_ablation}).
We then evaluate whether the gains over \rthree are statistically reliable and consistent across different candidate pools (Sections~\ref{sec:statistical_significance} and~\ref{sec:seed_robustness}).
We also compare \ours with the sorted-list selection strategy of MCS-SQL (Section~\ref{sec:mcs_comparison}).
Finally, we evaluate the robustness of question-based memory retrieval when the wording of a question changes (Section~\ref{sec:drspider_robustness}).

\subsection{Effect of Pointwise Tie-Breaking}
\label{sec:tiebreak_ablation}

We evaluate whether \ours depends on a separate pointwise reward model for tie-breaking.
We use BIRD-dev~\citep{li2024bird} with $n{=}8$ candidates generated by Agentar-Scale-SQL-Generation-32B~\citep{wang2025agentar}.
We compare the results with \rthree~\citep{r3-sql} under the same setting.
We evaluate three configurations of \ours.
The first configuration does not use tie-breaking.
The second configuration reuses Qwen3-Coder-30B-A3B-Instruct\footnote{\href{https://huggingface.co/Qwen/Qwen3-Coder-30B-A3B-Instruct}{Qwen3-Coder-30B-A3B-Instruct}} with a pointwise prompt.
The third configuration uses Contextual-RM-32B~\citep{contextual2025} as the pointwise tie-breaker.

Table~\ref{tab:tiebreak_ablation} shows that \ours remains competitive without a separate pointwise model.
Without tie-breaking, \ours achieves 72.23\% execution accuracy.
This result is higher than the 71.51\% of \rthree in the same setting.
Using Qwen3-Coder for tie-breaking improves the accuracy to 72.42\%.
This configuration reuses the same selector and does not require an additional reward model.
Using Contextual-RM-32B further improves the accuracy to 72.62\%.
These results show that the pointwise reward model is an optional accuracy enhancement.
It is not required by the core method.

\begin{table}[t]
    \centering
    \small
    \setlength{\tabcolsep}{6pt}
    \renewcommand{\arraystretch}{1.12}
    \begin{tabular}{@{}lc@{}}
        \toprule
        \textbf{Method} & \textbf{Acc.} \\
        \midrule
        \rthree & 71.51 \\
        \ours w/o tie-breaking & 72.23 \\
        \ours + Qwen3-Coder tie-break & 72.42 \\
        \ours + Contextual-RM tie-break & \textbf{72.62} \\
        \bottomrule
    \end{tabular}
    \caption{Tie-breaking ablation on BIRD-dev with $n{=}8$ candidates generated by Agentar-Scale-SQL-Generation-32B. Qwen3-Coder denotes Qwen3-Coder-30B-A3B-Instruct. Contextual-RM denotes Contextual-RM-32B. We report execution accuracy (\%).}
    \label{tab:tiebreak_ablation}
\end{table}

\subsection{Statistical Reliability of MAP-SQL improvements}
\label{sec:statistical_significance}

We evaluate whether the accuracy gains over \rthree are reliable at the question level.
We use the $n{=}32$ settings from our main experiments.
The evaluation covers BIRD-dev, Spider-test~\citep{yu2018spider}, and EHRSQL~\citep{lee2022ehrsql}.
The candidate pools are generated by Agentar and Arctic-Text2SQL-R1-7B~\citep{arctic2025text2sql}.
Both methods are evaluated on the same candidate pool.
We use a paired bootstrap to estimate a 95\% confidence interval for the difference in execution accuracy.
We also use the exact McNemar test to compare questions for which the two methods have different correctness outcomes.

Table~\ref{tab:statistical_significance} shows significant improvements in both BIRD settings and in Spider.
Their confidence intervals exclude zero.
Their exact McNemar test $p$-values are also below 0.05.
The EHRSQL confidence interval includes zero.
Its $p$-value is 0.125.
We therefore treat the EHRSQL improvement from this candidate pool as inconclusive.
The other three settings provide statistical support for the improvement of \ours over \rthree.

\begin{table*}[t]
\centering
\small
\setlength{\tabcolsep}{3pt}
\renewcommand{\arraystretch}{1.15}
\resizebox{\textwidth}{!}{%
\begin{tabular}{@{}llcccc@{}}
\toprule
\textbf{Dataset}
& \textbf{Generator}
& \textbf{\rthree Acc.}
& \textbf{\ours Acc.}
& \textbf{95\% CI for $\Delta$ Acc.}
& \textbf{Exact McNemar $p$} \\
\midrule
BIRD-dev
& Agentar-Scale-SQL-Generation-32B
& 71.97
& \textbf{73.08}
& [0.20, 2.02]
& 0.021 \\

BIRD-dev
& Arctic-Text2SQL-R1-7B
& 69.56
& \textbf{72.62}
& [1.76, 4.37]
& $<0.001$ \\

Spider-test
& Arctic-Text2SQL-R1-7B
& 86.82
& \textbf{87.59}
& [0.14, 1.39]
& 0.023 \\

EHRSQL
& Arctic-Text2SQL-R1-7B
& 44.03
& \textbf{44.71}
& [-0.06, 1.43]
& 0.125 \\
\bottomrule
\end{tabular}
}
\caption{Statistical significance of \ours over \rthree with $n{=}32$ candidates. We report paired bootstrap 95\% confidence intervals for the execution accuracy difference and exact McNemar test $p$-values.}
\label{tab:statistical_significance}
\end{table*}

\subsection{Robustness Across Randomly Generated Candidate Pools}
\label{sec:seed_robustness}

The preceding analysis evaluates both methods on a fixed candidate pool.
We further evaluate whether the comparison with \rthree depends on a particular candidate pool.
We generate three candidate pools with random seeds 42, 43, and 44.
Each pool contains $n{=}32$ candidates.
Both selectors are evaluated on every pool.
We report the mean execution accuracy and the sample standard deviation across the three pools.

Table~\ref{tab:seed_robustness} shows that \ours has higher mean execution accuracy in every dataset and generator setting.
This pattern also holds for EHRSQL despite the inconclusive result from the single candidate pool in Table~\ref{tab:statistical_significance}.
These results show that the improvement is consistent across the three generated candidate pools.

\begin{table}[t]
\centering
\small
\setlength{\tabcolsep}{4pt}
\renewcommand{\arraystretch}{1.12}
\resizebox{\columnwidth}{!}{%
\begin{tabular}{@{}llcc@{}}
\toprule
\textbf{Dataset}
& \textbf{Generator}
& \textbf{\rthree}
& \textbf{\ours} \\
\midrule
BIRD-dev
& Agentar-32B
& 71.82 $\pm$ 0.16
& \textbf{73.01 $\pm$ 0.36} \\

BIRD-dev
& Arctic-R1-7B
& 69.73 $\pm$ 0.20
& \textbf{72.34 $\pm$ 0.25} \\

Spider-test
& Arctic-R1-7B
& 86.92 $\pm$ 0.10
& \textbf{87.89 $\pm$ 0.26} \\

EHRSQL
& Arctic-R1-7B
& 43.86 $\pm$ 0.45
& \textbf{45.28 $\pm$ 0.52} \\
\bottomrule
\end{tabular}
}
\caption{Robustness across candidate pools generated with seeds 42, 43, and 44 for $n{=}32$. We report mean execution accuracy and sample standard deviation.}
\label{tab:seed_robustness}
\end{table}

\subsection{Comparison with MCS-SQL}
\label{sec:mcs_comparison}

We compare our group-based permutation strategy with the sorted-list strategy of MCS-SQL~\citep{lee2024mcs}.
The official implementation of MCS-SQL was not publicly available when we conducted this experiment.
We therefore reproduce its multiple-choice selection strategy.
We use Qwen3-Coder-30B-A3B-Instruct as the selector for both methods.
Both methods also use the same $n{=}8$ candidate pools.
This setup isolates the difference between the two selection strategies.

Table~\ref{tab:mcs_comparison} shows that \ours achieves higher execution accuracy in all four settings.
The improvement appears with both generators on BIRD-dev.
The same pattern also appears on Spider-test and EHRSQL.
These results show that group-based permutation performs better than the reproduced sorted-list strategy under the same selector and candidate pools.

\begin{table}[t]
\centering
\small
\setlength{\tabcolsep}{3pt}
\renewcommand{\arraystretch}{1.12}
\resizebox{\columnwidth}{!}{%
\begin{tabular}{@{}lcccc@{}}
\toprule
\textbf{Method}
& \shortstack{\textbf{BIRD-dev} \\ \textbf{Agentar}}
& \shortstack{\textbf{BIRD-dev} \\ \textbf{Arctic}}
& \shortstack{\textbf{Spider-test} \\ \textbf{Arctic}}
& \shortstack{\textbf{EHRSQL} \\ \textbf{Arctic}} \\
\midrule
MCS-SQL
& 71.51
& 70.47
& 86.97
& 39.25 \\

\ours
& \textbf{72.62}
& \textbf{72.16}
& \textbf{87.49}
& \textbf{39.93} \\
\bottomrule
\end{tabular}
}
\caption{Comparison with the MCS-SQL~\cite{lee2024mcs} sorted-list strategy using $n{=}8$ candidates. We report execution accuracy (\%).}
\label{tab:mcs_comparison}
\end{table}

\subsection{Robustness on Dr.Spider}
\label{sec:drspider_robustness}

We evaluate whether question-based memory retrieval remains effective when the wording of a question changes.
Our retriever uses bge-m3~\citep{chen2024bge} to retrieve memories based on question similarity.
We use the nine natural-language question perturbations from Dr.Spider~\citep{chang2023drspider}.
These perturbations change how the original questions are expressed while preserving their intended meaning.
All methods use $n{=}8$ candidates generated by Arctic-Text2SQL-R1-7B.
We report accuracy before perturbation as Pre.
We report accuracy after perturbation as Post.
Drop is calculated as Pre minus Post.

Table~\ref{tab:drspider_robustness} shows that listwise selection with memory achieves the highest Post accuracy of 77.80\%.
It also has the smallest Drop at 12.72 points.
Listwise selection without memory reaches 77.06\% after perturbation.
Its Drop is 13.13 points.
These results show that the retrieved memories remain useful when questions are expressed with different wording.
They also show that question-based retrieval remains robust when the intended meaning is preserved but the wording changes.

\begin{table}[t]
\centering
\small
\setlength{\tabcolsep}{4pt}
\renewcommand{\arraystretch}{1.12}
\resizebox{\columnwidth}{!}{%
\begin{tabular}{@{}lccc@{}}
\toprule
\textbf{Method}
& \textbf{Pre $\uparrow$}
& \textbf{Post $\uparrow$}
& \textbf{Drop $\downarrow$} \\
\midrule
greedy
& 89.54
& 76.22
& 13.32 \\

majority voting
& 89.23
& 76.43
& 12.80 \\

pointwise
& 89.93
& 76.66
& 13.27 \\

pairwise
& 89.56
& 76.13
& 13.43 \\

listwise w/o Memory
& 90.13
& 77.06
& 13.13 \\

listwise w/ Memory (ours)
& \textbf{90.52}
& \textbf{77.80}
& \textbf{12.72} \\
\bottomrule
\end{tabular}
}
\caption{Average execution accuracy (\%) across the nine natural-language question perturbations of Dr.Spider with $n{=}8$ candidates generated by Arctic-Text2SQL-R1-7B. Drop is calculated as Pre minus Post.}
\label{tab:drspider_robustness}
\end{table}

Taken together, these experiments support the reliability and robustness of the reported improvements.
The gains over \rthree receive statistical support in three of the four fixed-pool settings and remain consistent across the three generated candidate pools.
The results also show that \ours remains competitive without a separate pointwise reward model and outperforms the reproduced MCS-SQL strategy in all four evaluated settings.
Finally, the Dr.Spider results show that retrieved memories remain useful when questions are expressed with different wording.

\section{Additional Examples and Prompts}
\label{sec:additional_examples}

\subsection{Memory Examples and Generation Prompt}
\label{sec:memory_examples}

Our memory design follows the Encoding, Translating, and Decoding decomposition of \citet{deng2022recent}.
Figure~\ref{fig:memory_example} shows an example memory generated from a training question and its gold SQL.
Figures~\ref{fig:memory_prompt} and~\ref{fig:qual} show the generation prompt and an example of how the retrieved memory guides listwise selection.

\begin{figure*}[t]
    \small
    \centering
    \begin{tcolorbox}[
        title=\textbf{Example of Retrieved Memory Used for Listwise Selection},
        left=3pt, right=3pt, top=3pt, bottom=3pt,
        enhanced
    ]

\textbf{Question:} \\
\texttt{How many athletes from region 151 have won a medal?}

\vspace{0.5em}
\textbf{Database Schema (relevant subset):} \\
\texttt{
CREATE TABLE competitor\_event ( \\
\ \ event\_id INTEGER, \\
\ \ competitor\_id INTEGER, \\
\ \ medal\_id INTEGER, \\
\ \ FOREIGN KEY (competitor\_id) REFERENCES games\_competitor(id) \\
); \\
\\
CREATE TABLE games\_competitor ( \\
\ \ id INTEGER, \\
\ \ person\_id INTEGER, \\
\ \ PRIMARY KEY (id), \\
\ \ FOREIGN KEY (person\_id) REFERENCES person(id) \\
); \\
\\
CREATE TABLE person\_region ( \\
\ \ person\_id INTEGER, \\
\ \ region\_id INTEGER \\
);
}

\vspace{0.5em}
\textbf{Specification (JSON):}
\begin{verbatim}
{
  "schema_grounding": {
    "mappings": [
      "athletes -> person_region.person_id",
      "region 151 -> person_region.region_id = 151",
      "won a medal -> competitor_event.medal_id != 4"
    ]
  },
  "join_path": {
    "paths": [
      "competitor_event --(competitor_id)--> games_competitor",
      "games_competitor --(person_id)--> person_region"
    ],
    "join_types": ["INNER", "INNER"]
  },
  "filter_semantics": {
    "conditions": ["person_region.region_id = 151", "competitor_event.medal_id != 4"]
  },
  "aggregation": {
    "group_by_columns": [],
    "aggregate_functions": "COUNT(person_region.person_id)"],
    "distinct_within_aggregate": "false"
  },
  "output_form": {
    "result_shape": "single_value",
    "return_columns": ["COUNT(person_region.person_id)"],
    "distinct": "false",
    "format": "number"
  },
  "query_constraints": {
    "avoid_unnecessary_distinct": "true", "avoid_extra_join": "true",
    "avoid_wrong_aggregation_column": "true", "avoid_extra_filter": "true",
    "prefer_schema_defined_columns": "true"
  },
  "extra_keywords": []
}
\end{verbatim}

    \end{tcolorbox}
    \caption{Example of an LLM-generated memory derived from a training question--SQL pair.}
    \label{fig:memory_example}
\end{figure*}
\begin{figure*}[t]
    \small
    \centering
    \begin{tcolorbox}[
        title=\textbf{Prompt Template for Memory Generation},
        left=3pt, right=3pt, top=3pt, bottom=3pt,
        enhanced
    ]

\begin{verbatim}
You are an expert in Text-to-SQL reasoning.
Analyze the given natural language question, database schema, and correct SQL query.
Generate a structured specification memory that helps an LLM select the correct SQL query 
among multiple candidates.

--------------------------------------------------
INPUT

[Question]
{question}

[Database Schema]
{schema}

[Correct SQL Query]
{gold_sql}

[Detected Extra SQLite Constructs in Gold SQL]
The following keywords / functions were found in the correct SQL.
You MUST populate "extra_keywords" with one entry per detected keyword.
Detected: {detected_extra_keywords}

--------------------------------------------------
OUTPUT FORMAT

Return a single flat JSON object.

IMPORTANT RULES:
- Include a key ONLY when its corresponding SQL construct actually appears in the correct SQL. 
  Omit keys entirely when they are not relevant.
- "extra_keywords" is ALWAYS included. Use [] when no extra keywords are detected.

{
  "schema_grounding": {
    "mappings": ["question phrase -> table.column", ...]
  },
  "join_path": {
    "paths": ["tableA --(fk)--> tableB"],
    "join_types": ["INNER | LEFT | CROSS | ..."]
  },
  "filter_semantics": {
    "conditions": ["column op value, e.g. release_year = 1945, price BETWEEN 10 AND 50"],
  },
  "aggregation": {
    "group_by_columns": ["col1"],
    "aggregate_functions": ["COUNT(*)", "SUM(amount)"],
    "having_condition": "e.g. COUNT(*) > 3",
    "distinct_within_aggregate": "true | false"
  },
  "ordering_and_scope": {
    "order_by": ["col ASC | DESC"], "limit": "N", "offset": "M"
  },
  "conditional_and_null": {
    "construct": "CASE | IIF | COALESCE | IFNULL | NULLIF", ...
  },
  "output_form": {
    "result_shape": "single_value | single_row | multiple_rows",
    "format": "number | year | text | id | boolean | percentage | custom", ...
  },
  "query_constraints": {
    "avoid_unnecessary_distinct": "true | false", ...
  },
  "extra_keywords": [{"keyword": "X", "usage": "one sentence on how it is used in this query"}, ...
  ]
}
\end{verbatim}
    \end{tcolorbox}
    \caption{Prompt template used for memory generation.}
    \label{fig:memory_prompt}
\end{figure*}
\tcbset{qualbox/.style={enhanced,breakable,boxsep=1pt,left=3pt,right=3pt, top=2pt,bottom=2pt,fonttitle=\bfseries\small}}
\begin{figure*}[t]
    \footnotesize
    \centering
    \begin{tcolorbox}[qualbox,title=Qualitative Example --- memory resolves three near-tied candidates for one question]
\textbf{Question} (\textsc{codebase\_community}): \texttt{Among posts by Harvey Motulsky and Noah Snyder, which one has higher popularity?}\\[1pt]
\textbf{Retrieved memory} (most relevant of the $k$ retrieved; \colorbox{green!18}{green} = decisive):
\begin{lstlisting}[style=memjson]
{
  "schema_grounding": { "mappings": ["reviewer -> ProductReview.ReviewerName"] },
  "output_form": { "result_shape": "single_row",
      "return_columns": [(*\colorbox{green!18}{\texttt{"ProductReview.ReviewerName"}}*)], "format": "text" },
  "query_constraints": { "avoid_extra_filter": "true", "prefer_schema_defined_columns": "true" }
}
\end{lstlisting}
\textbf{Candidates} (all execute; they disagree on \emph{what to return})
\begin{description}\setlength\itemsep{1pt}\setlength\parskip{0pt}
\item[(A)] \texttt{SELECT \colorbox{green!18}{u.DisplayName} ... GROUP BY u.DisplayName ORDER BY SUM(p.ViewCount) DESC;}\hfill$\Rightarrow$ \texttt{Harvey Motulsky}
\item[(B)] \texttt{SELECT \colorbox{red!12}{p.Title} ... ORDER BY p.ViewCount DESC LIMIT 1;}\hfill$\Rightarrow$ \texttt{"Power of Holm's ..."}
\item[(C)] \texttt{SELECT u.DisplayName, \colorbox{red!12}{SUM(p.ViewCount)} ... GROUP BY u.DisplayName ...;}\hfill$\Rightarrow$ \texttt{Harvey Motulsky | 23065}
\end{description}
\textbf{Listwise output:} \texttt{(A) $>$ (C) $>$ (B) $> \cdots$}\quad (A selected)
\tcblower
\centering
\resizebox{\linewidth}{!}{%
\colorbox{green!15}{\texttt{memory.output\_form}: return only the queried entity's name} $\;\to\;$
\colorbox{green!18}{(A) \texttt{DisplayName}}~\cmark\ a person \quad
\colorbox{red!12}{(B) \texttt{Title}}~\xmark\ a post \quad
\colorbox{red!12}{(C) \texttt{+SUM}}~\xmark\ extra column}\\[3pt]
\fcolorbox{black!40}{yellow!20}{first-stage rank \textbf{18}\,$\Longrightarrow$\,final rank \textbf{1}}
    \end{tcolorbox}
    \caption{\textbf{Memory resolves three near-tied candidates.} For a single question, three candidates all execute and return a plausible value, so execution alone cannot rank them: (A)~returns a display name (\texttt{Harvey Motulsky}), (B)~returns a post title (\texttt{"Power of Holm's ..."}), and (C)~returns a name together with its view count (\texttt{Harvey Motulsky, 23065}). The question asks which of the two \emph{people} is more popular, so the answer should be a single person's name. The retrieved memory's \texttt{output\_form} field prescribes returning exactly the queried entity's name column---mirroring its twin \emph{``Which reviewer\,...\,$\to$\,ReviewerName''}---which selects~(A) over the post-valued~(B) and the over-specified~(C); conditioning the listwise selector on the memory promotes~(A) from rank $18/32$ to the top.}
    \label{fig:qual}
\end{figure*}

\subsection{Selection Prompts}
\label{sec:selection_prompts}

We provide the full prompts used for listwise and pairwise selection.
Figure~\ref{fig:listwise_prompt_template} shows the prompt used by our listwise selector.
Figure~\ref{fig:pairwise_prompt_template} shows the pairwise prompt used for the CHASE-SQL-style baseline~\citep{pourreza2025chase}.

\begin{figure*}[t]
    \small
    \centering
    \begin{tcolorbox}[
        title=\textbf{Full Prompt Template for Listwise SQL Selection},
        left=3pt, right=3pt, top=3pt, bottom=3pt,
        enhanced
    ]

\textbf{Prompt template:}
{\small
\begin{verbatim}
Given the db info and question, there are {num_candidates} candidate queries.
Compare all candidates, analyze the differences in their queries and execution results.
{score_instruction}
Based on the original question and the provided database info, rank them based on their correctness.

**************************
Database Schema
{schema}
**************************
Question: {question}
Evidence: {hint}
**************************
Candidate A
```sql
{candidate_A_sql}
```
Execution result:
{candidate_A_execution_result}
{optional_consensus_or_judge_score_A}
**************************
Candidate B
```sql
{candidate_B_sql}
```
Execution result:
{candidate_B_execution_result}
{optional_consensus_or_judge_score_B}
**************************
...
Candidate {last_label}
```sql
{candidate_last_sql}
```
Execution result:
{candidate_last_execution_result}
{optional_consensus_or_judge_score_last}
**************************
**************************

Rank the passages in descending order of correctness to the query.
The most correct query identifier should be placed first.
Output ONLY letter identifiers in the format of, e.g., [A] > [B] > ... .
Do NOT say any word or explain.

Ranking:
\end{verbatim}
}

    \end{tcolorbox}
    \caption{Full prompt template used by the listwise SQL reranker. Candidate SQLs, execution results, optional scores, and retrieved memory sections are filled at runtime before the model is asked to output a ranking such as \texttt{[A] > [B] > ...}.}
    \label{fig:listwise_prompt_template}
\end{figure*}

\begin{figure*}[t]
    \small
    \centering
    \begin{tcolorbox}[
        title=\textbf{Full Prompt Template for Pairwise SQL Selection},
        left=3pt, right=3pt, top=3pt, bottom=3pt
    ]

\textbf{Prompt template:}
{\small
\begin{verbatim}
Given the DB info and question, there are two candidate queries.
There is correct one and incorrect one, compare the two candidate answers,
analyze the differences of the query and the result.
{score_instruction}
Based on the original question and the provided database info, choose the correct one.

**************************
Database Schema
{schema}
**************************
Question: {question}
Evidence: {hint}
**************************
Candidate A
```sql
{candidate_a}
```
Execution result:
{result_a}{score_text_a}{judge_text_a}
**************************
Candidate B
```sql
{candidate_b}
```
Execution result:
{result_b}{score_text_b}{judge_text_b}
**************************

Output ONLY "A" or "B" to indicate the correct candidates.
Do NOT say anything or explain.

Correct Answer:
\end{verbatim}
}

    \end{tcolorbox}
    \caption{Full prompt template used by the pairwise SQL reranker. For each non-identical pair of candidate execution results, the model compares two SQL queries and returns exactly \texttt{A} or \texttt{B}.}
    \label{fig:pairwise_prompt_template}
\end{figure*}

\section{Usage of AI Assistants}
We utilized ChatGPT to improve the clarity and grammatical accuracy of the writing. It provided suggestions for rephrasing sentences and correcting grammatical errors to make the text flow more naturally.

\end{document}